\documentclass[twoside]{article}
\usepackage[latin1]{inputenc}
\usepackage{t1enc}
\usepackage{a4}
\usepackage{tabularx}
\usepackage{psfig}

\usepackage[english]{babel}

\def\bref{\vspace{4pt}\noindent\hangindent=10mm}

\newcommand{\vm}{V_{\rm max}}
\newcommand{\oii}{[O\,{\small II}]}
\newcommand{\oiii}{[O\,{\small III}]}
\newcommand{\rd}{r_{\rm d}}

\begin{document}

\setcounter{figure}{0}
\setcounter{section}{0}
\setcounter{equation}{0}

\begin{center}
{\Large\bf
The Evolution of Field Spiral Galaxies\\[0.2cm]
over the Past 8\,Gyr}\\[0.7cm]

Asmus B\"ohm and Bodo L. Ziegler\\[0.17cm]
University Observatory G\"ottingen \\
Geismarlandstr.~11, 37083 G\"ottingen, Germany \\
boehm@uni-sw.gwdg.de
\end{center}

\vspace{0.5cm}

\begin{abstract}
\noindent{\it
We have performed a large observing campaign of intermediate--redshift 
disk galaxies including multi--object spectroscopy with
the FORS instruments of the Very Large Telescope
and imaging with the Advanced Camera for Surveys onboard the Hubble Space
Telescope. 
Our data set comprises 113 late--type galaxies in the redshift 
range $0.1<z<1.0$ and thereby
probes galaxy evolution over more than half the age of the universe.
Spatially resolved rotation curves have been extracted and fitted with
synthetic velocity fields that account for geometric distortions and
blurring effects.
With these models,
the intrinsic maximum rotation velocity $V_{\rm max}$ was derived for
73 spirals within the field--of--view of the ACS images. 
Combined with the structural parameters from 
two-dimensional surface brightness profile fitting, the scaling relations
(e.g., the Tully--Fisher Relation) at intermediate redshift 
were constructed. 
The evolution of these relations offers powerful tests of the predictions
of simulations within the Cold Dark Matter 
hierarchical scenario.

By comparing our sample to the Tully--Fisher Relation of local spiral galaxies,
we find evidence for a differential luminosity evolution:
the massive distant galaxies are of comparable luminosity as 
their present-day counterparts,
while the distant low--mass spirals are brighter than locally by up to
$>$2$^m$ in rest--frame $B$. Numerous tests applied to the data 
confirm that this trend is
unlikely to arise from an observational bias or systematic errors.
Discrepancies between several previous studies could be explained
as a combination of selection effects and small number statistics
on the basis of such a mass--dependent luminosity evolution.
On the other hand, this evolution would be at variance with 
the predictions from numerical simulations. 
For a given $\vm$, the disks of the distant 
galaxies are slightly smaller than those of their local counterparts, 
as expected for a hierarchical structure growth. 
Hence, the discrepancy between the observations and theoretical predictions
is limited to the properties of the stellar populations.
A possible explanation could be the suppression of star formation
in low--mass disks which is not
yet properly implemented in models of galaxy evolution.
}
\end{abstract}

\section{\label{intro}Introduction}
Within the last few years, our knowlegde of the basic parameters which
determine the past, present and future of the universe has improved
significantly. Thanks to the combined results from studies of
the Cosmic Microwave Backgrund, the Large Scale Structure, Big Bang
Nucleosynthesis and distant supernovae, 
we now have strong evidence for a flat metric of spacetime
(Spergel et al.~2003 and references therein).
According to the observations, 73\% of the mean density of the universe
originate from Dark Energy, 23\% are contributed by Cold Dark 
Matter and only 4\% by ``ordinary'' baryonic matter. In such a cosmology,
structure growth proceeds hierarchically, with small structures forming
first in the early cosmic stages, followed by the successive build-up
of larger structures via merger and accretion events.

Although the constituents of the Dark Energy and Dark Matter remain unknown,
the 
$\Lambda$CDM or ``concordance'' cosmology has been a very successful tool
for the interpretation of structures on Mpc scales 
and beyond (e.g., Peacock 2003).
On scales of individual galaxies, however, several discrepancies between 
observational results and theoretical predictions have been found,
a prominent of which is the ``angular momentum problem''. 
This term depicts the loss of angular momentum of the baryons to the surrounding
DM halo, resulting in galactic disks within numerical simulations
which are smaller than observed (e.g., Navarro \& White 1994),
however more recent studies made progress in this respect
(e.g., Governato et al.~2004).
Aiming at a quantitative test of the hierarchical scenario at the scale
of individual galaxies, we performed an observational study 
which covers a significant fraction of the Hubble time.

For this purpose, we utilised scaling relations 
like the Tully--Fisher relation (TFR, Tully \& Fisher 1977) between the 
luminosity $L$ and the maximum rotation velocity $\vm$ of spiral
galaxies.
Basically, this correlation can be understood as a combination of
the virial theorem and the rotational stabilisation of late--type galaxies.
By comparing local and distant spirals of a given $\vm$,
the luminosity evolution within the look--back time can be determined. 
Since the maximum rotation velocity
is a measure for the total (virial) mass of a disk galaxy 
($\vm^2 \propto M_{\rm vir}$, e.g.~van den Bosch 2002), 
the TF analysis relates the evolution of stellar population properties
to the depth of the gravitational potential well.   

Numerical simulations within CDM-do\-mi\-na\-ted 
cosmologies have been successfully used to reproduce the observed slope 
of the local TFR, whereas the zero points were offset due to dark 
halos with too high concentrations (e.g., Steinmetz \& Navarro 1999). 
The TFR slope was predicted to remain constant with cosmic look--back time 
in such $N$-body simulations; nevertheless the modelling of realistic stellar
populations at sufficient resolution remains a challenge~--- typically,
the masses of individual particles are of the order of 
$10^5 M_\odot$ \dots $10^6 M_\odot$.
Other theoretical approaches focussed more on the chemo--spectrophotometric
aspects of disk galaxy evolution. 
For example, Boissier \& Prantzos (2001) calibrated their models 
to reproduce the observed colors of local spirals. 
Compared to these, the authors predicted
higher luminosities for massive disks and lower luminosites for low--mass disks at
redshifts $z>0.4$.
A similiar evolution was found by Ferreras \& Silk (2001).
By modelling the mass--dependent chemical enrichment history 
of disk galaxies with the local TFR as a constraint, the authors found a 
TFR slope that increases with  look--back time 
(i.e., for a parameterisation $L \propto V_{\rm max}^\alpha$, $\alpha$ 
increases with redshift).
 
In the last decade, many observational studies of the \emph{local} TFR 
have produced very large samples with $N_{\rm obj} \approx 1000$ 
(e.g.  Haynes et al.~1999), not only to derive 
the slope and scatter with high accuracy, but also to map the peculiar 
velocity field out to $cz \approx 15000$\,km\,s$^{-1}$ 
(e.g.  Mathewson \& Ford 1996). 
Other groups used spirals, partly with cepheid--calibrated distances, to 
measure the Hubble constant.
For example, Sakai et al.~(2000) derived a value of 
$H_0 = (71 \pm 4)$\,km\,s$^{-1}$\,Mpc$^{-1}$ with this method.

At higher redshifts, robust measurements of rotation velocities 
become increasingly difficult, 
which is mainly for two reasons. Firstly, because the objects are very faint.
Given a redshift of $z=0.5$, the surface brightness 
at galactocentric radii of $\sim$\,3\,$\rd$~--- where the regime
of constant rotation velocity is reached~--- is typically
$\mu_B \approx 27$\,mag\,arcsec$^{-2}$. Spatially resolved spectroscopy
at this level has become feasible just with the generation of
10m-class telescopes.
The second difficulty coming into play arises from the small apparent sizes
of the galaxies, this issue will be described in detail in
Sect.~\ref{anakin}.

A number of samples with 10-20 objects in the regime
$0.25< \langle z \rangle <0.5$ have been observed  
to estimate a possible evolution in luminosity by comparison to the local TFR. 
The results of these studies were quite discrepant:
e.g. Vogt et al.~(1996, 1997) found only a modest increase in luminosity
of  $\Delta M_B \approx -0.5^m$, whereas Simard \& Pritchet (1998) and
Rix et al.~(1997) derived a much stronger brightening with
$\Delta M_B \approx -2.0^m$. A more recent study of 19 field spirals by
Milvang-Jensen et al.~(2003) yielded a value of
$\Delta M_B \approx -0.5^m$ and showed evidence for an increase of this
brightening with redshift. 

It seems likely that some of these results are affected by 
the selection criteria. For example, Rix et al.~selected blue colors with
$(B-R)_{\rm obs}<1.2^m$, Simard \& Pritchet strong [O\,{\small II}] 
emission with equivalent widths $>$20\,\AA, while Vogt et al.~partly chose
large disks with $r_{\rm d}>3$\,kpc. 
The two former criteria prefer late--type spirals, whereas the latter
criterion leads to the overrepresentation of large, early--type spirals.
Additionally, due to the small samples, all these studies had to assume that 
the local TFR slope holds valid at intermediate redshifts~---
we will adress this topic again in Sect.~\ref{dis}. 

Based on a larger data set from the DEEP Groth Strip Survey 
(Koo 2001) with $N \approx 100$ spirals in the range 
$0.2~<~z~<~1.3$, Vogt (2001) found a constant TFR slope and only a 
very small rest--frame $B$-band brightening of $\le$\,0.2$^m$.
On the other hand, in a more recent study based on the same survey,
an evolution of the luminosity--metallicity relation 
both in slope and zero point was observed (Kobulnicky et al.~2003).
The authors argued that low--luminosity galaxies
probably have undergone a decrease in luminosity combined with  
an increase in metallicity during the last $\sim$\,8\,Gyr. 

Throughout this article,
we will assume the concordance cosmology with
$\Omega_{\rm m}$ = 0.3, $\Omega_\Lambda$ = 0.7 and 
$H_0$ = 70\,km\,s$^{-1}$\,Mpc$^{-1}$.

\section{\label{sel}Sample Selection \& Observations}

The sample described here has been selected within the FORS Deep
Field (FDF, see Heidt et al.~2003), an $UBgRIJK$ photometric survey
covering a sky area of $\sim$\,6\,$\times$\,6\,arcmin$^2$ near
the southern Galactic pole. The imaging was  
performed with the Very Large Telescope (optical bands) and the
New Technology Telescope (Near Infrared bands).
Based on a catalogue with spectral types and photometric redshift 
estimates (Bender et al.~2001), we chose objects for follow--up spectroscopy
which satisfied the following criteria: 
{\it 1)} late--type Spectral Energy Distribution, 
i.e., galaxies with emission lines,
{\it 2)} total apparent $R$-band magnitude $R\le23^m$,
{\it 3)} photometric redshift
$z_{\rm phot}\le1.2$ to ensure that at least
the \oii 3727 doublet falls within the wavelength range of the spectra,
{\it 4)} disk inclination angle $i \ge 40^\circ$ and
{\it 5)} deviation between slit direction and apparent major
axis of $\delta \le 15^\circ$. The two latter constraints were chosen to
limit the geometric distortions of the observed rotation curves.
For some objects, however, these limits had to be exceeded during
the construction of the spectroscopic setups. 

After a pilot observation in 1999, the spectroscopy was performed in
2000 and 2001 using the FORS1 \& 2 instruments of the VLT in
multi--object spectroscopy mode with a total integration time of 2.5\,hrs
per setup. Using the  grism \texttt{600R}, a spectral
resolution of $R\approx1200$ was achieved with a spectral scale of
1.07\,\AA/pix and a spatial scale of 0.2\,arcsec/pix.
The seeing ranged between 0.4 and 1.0\,arcsec  with a median of
0.74\,arcsec. In total, 129 late--type galaxies were observed.

For an accurate derivation of the galaxies' structural parameters, like disk 
inclination, scale length etc., we also took Hubble Space Telescope
images of the FDF with the Advanced Camera for Surveys 
using the F814W filter. To cover the complete FDF area, a 
2~$\times$~2 mosaic was observed. 

\section{\label{anaspec}Spectrophotometric Analysis}
 
The spectra of 113 galaxies were reliable for redshift determination. 
Out of these,
73 objects eventually yielded maximum rotation velocities
(see next section) and were covered by the HST/ACS imaging;
these objects will be referred to as the FDFTF sample in the following.
They span the 
redshift range $0.09<z<0.97$ with a median of
$\langle z \rangle = 0.45$ corresponding to look--back times 
$1.2\,{\rm Gyr} < t_1 < 7.6\,{\rm Gyr}$ with 
$\langle t_1 \rangle = 4.7$\,Gyr. 
This data set covers all spectrophotometric
types from very early--type spirals (Sa or $T=1$) to very late--type galaxies 
(Sdm/Im or $8 \le T \le 10$).

An analysis of the galaxies' surface brightness profile 
profiles was conducted with the
GALFIT package (Peng et al.~2002). To fit the
disk component, an exponential profile was used, while a potential bulge 
was approximated with a S\'ersic profile. In the case of 13 FDFTF galaxies,
the fit residual images and large fit errors indicated an irregular
component that could not be approximated properly with a S\'ersic law.
The bulge--to--total  ratios of these galaxies were assumed to be undefined.
The $B/T$ ratios of the other 60 FDFTF galaxies ($0\le B/T \le 0.53$ with 
$\langle B/T \rangle = 0.04$)
confirm that the vast majority of these galaxies 
are disk-dominated.

Total apparent magnitudes were determined using the \texttt{mag\_auto} algorithm
of the Source Extractor package
(Bertin \& Arnouts 1996). For the computation of absolute $B$-band magnitudes
$M_B$, we used the filter which, depending on the redshift of a given object, 
best matched the rest--frame $B$-band. 
For galaxies at $z \le 0.25$, $0.25 < z \le 0.55$, $0.55 < z \le 0.85$ and
$z>0.85$, we thus utilised the $B$, $g$, $R$ and $I$ magnitudes, respectively.
Thanks to this strategy, the $k$-correction uncertainties $\sigma_k$ 
---~usually a substantial source of error to the luminosities 
of distant galaxies~---
are smaller than 0.1$^m$ for all types and redshifts in our sample. 
For the correction of intrinsic dust absorption, we followed the approach
of Tully \& Fouqu\'e (1985) assuming a face--on ($i=0^\circ$)
extinction of $A_B=0.27^m$. The absolute magnitudes of the FDFTF galaxies
computed this way span the range $-18.0^m \ge M_B \ge -22.7^m$.

The spectra of 12 objects in our sample cover a wavelength range
that simultaneously shows emission in \oii3727, H$\beta$, 
\oiii 4959 and \oiii 5007 at sufficient signal--to--noise to determine 
the equivalent widths. These lines can be used to estimate the gas-phase
metallicity. We adopted the analytical expressions given by McGaugh  (1991)
to compute the abundances O/H from the $R_{23}$ and $O_{32}$ parameters.
Since all the galaxies have $M_B<-18^m$, we assumed that they fall on the
metal-rich branch of the $R_{23}$--O/H relation. 
The galaxies have abundances $8.37<\log ({\rm O/H})<8.94$.
We will use these estimates to investigate the luminosity--metallicity
relation in Sect.~\ref{scal}.

\section{\label{anakin}Derivation of $\mathbf \vm$}

We extracted spatially resolved rotation curves from the two--dimensional
spectra by
fitting Gaussians to the emission lines stepwise along the spatial axis.
Line fits at any projected radius 
which, compared to the instrumental broadening 
(FWHM$_{\rm ins} \approx 4.5$\,\AA), had very small 
(FWHM$_{\rm fit} < 2$\,\AA) or very large
(FWHM$_{\rm fit} > 12$\,\AA) line widths were assumed to be
noise and therefore neglected.

The analysis of spatially resolved rotation curves from optical
spectroscopy of \emph{local} spiral galaxies is relatively straightforward.
But in the case of distant galaxies with very small apparent sizes, 
the effect of the slit width
on the observed rotation velocities  $V_{\rm rot}(r)$ must be considered.
At redshift $z=0.5$, a scale length of 3\,kpc --- typical for an 
$L^{\ast}$ spiral --- corresponds
to $\sim$\,0.5\,arcsec only, which is half the slit width used in our 
observations. 
Any value of $V_{\rm rot}(r)$ is therefore an integration 
perpendicular to the spatial axis (slit direction), a phenomenon which 
is the optical equivalent to ``beam smearing'' in radio observations. 
The seeing has an additional blurring effect on the observed
rotation curves.
If not taken into account, these two phenomena would lead to an underestimation 
of the \emph{intrinsic} rotation velocities and, in particular,
the \emph{intrinsic} $\vm$.

We overcame this problem by generating synthetic rotation curves.
For the intrinsic rotational law, 
we used a simple shape
with a linear rise of $V_{\rm rot}(r)$ at small radii, turning over into
a region of constant rotation velocity where the Dark Matter halo dominates the
mass distribution. Alternatively, we also tested the so--called 
``Universal Rotation Curve'' shape (Persic et al.~1996), a parameterisation which
introduces a velocity gradient in the outer regions of the disk which
is positive for sub-$L^\ast$ objects and negative for objects much more luminous
than $L^\ast$. However, the results given here are
not sensitive to the form of the intrinsic rotational law~---
see B\"ohm et al.~(2004) for a detailed discussion of this topic~--- and we 
therefore only use $\vm$ values determined with an intrinsic 
``rise--turnover--flat'' shape here.

\begin{figure}[t]
\centerline{
\hspace{-1cm}
\psfig{file=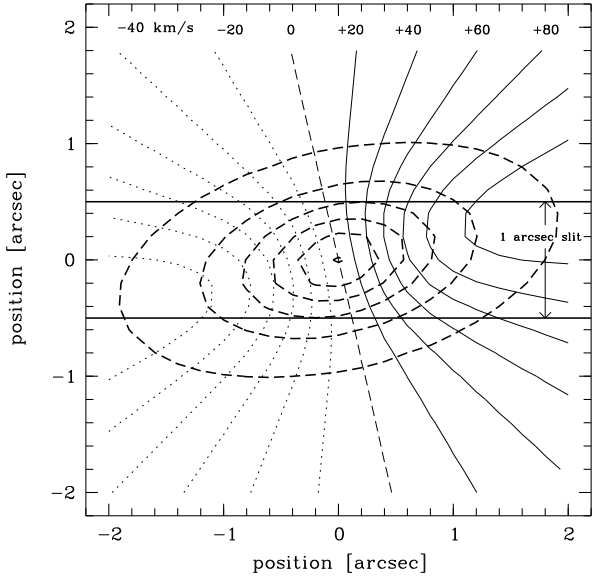,width=7.8cm,angle=-90}
}
\caption{\label{vf}
Example of a simulated rotation velocity field for an object from our
data set with a disk inclination $i=64^\circ$ and a misalignment angle
between apparent major axis and slit direction of
$\delta=+13^\circ$. 
The dashed ellipses denote the isophotes of the disk, with
the outermost corresponding to an $I$-band surface brightness of 
$\mu_I \approx25$\,mag\,arcsec$^{-2}$.
The curved dotted and solid lines correspond to line--of--sight
rotation velocities
ranging from $-$120\,km/s to $+$120\,km/s.
The two solid horizontal lines visualise the position of
the slit used for spectroscopy.
}
\end{figure}

\begin{figure}[t]
\centerline{
\psfig{file=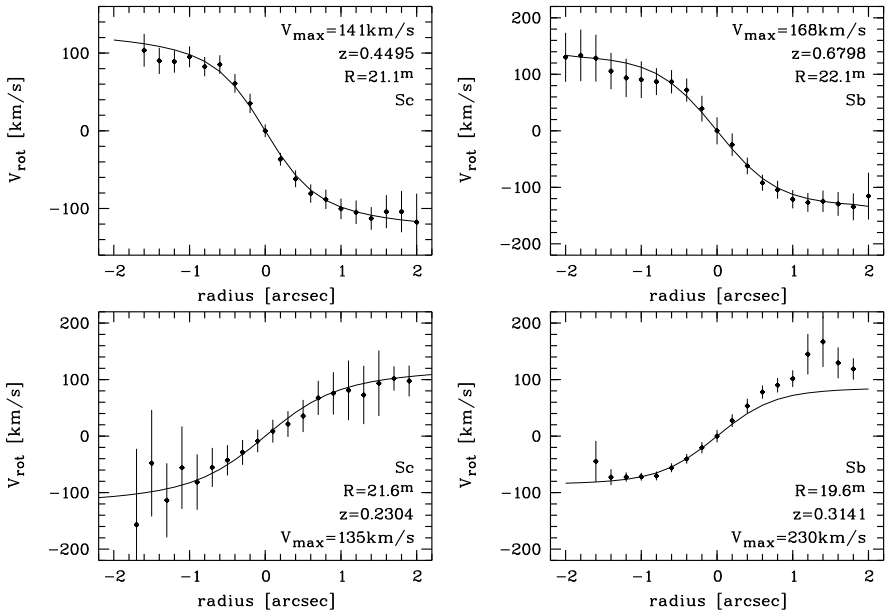,width=11.8cm,angle=-90}
}
\caption{\label{rcs}
Examples of rotation curves from our data set. The solid lines are the
synthetic rotation curves fitted to the observed rotation velocity as a
function of radius (solid symbols) used to derive the intrinsic maximum
rotation velocity. For each object the spectrophotometric type, total
apparent $R$\, magnitude, redshift and $\vm$ are given.
The two upper curves were classified as high quality data, the two
lower ones as low quality data due to the large measurement errors
(lower left) or an asymmetric shape (lower right).
}
\end{figure}

Given the observed inclination, position angle and scale length of an object,
the intrinsic rotation velocity field was constructed, Fig.~\ref{vf} 
shows an example.
In the next step of the simulation, the velocity field was weighted with
the surface brightness profile.
The effect of this was that, just like for the observed data,
brighter regions contributed stronger to the rotation
velocities in direction of dispersion than fainter regions
(the ``beam smearing'' effect). 
Following the weighting,
the velocity field was convolved with the Point Spread Function
to simulate the blurring due to seeing. 
Finally, a ``stripe'' was extracted from the
velocity field, with a position and width that corresponded to the slit
used during the observations, and integrated perpendicular to the spatial
axis. The results of the whole procedure was a 
synthetic rotation curve which introduced the same 
geometric and blurring effects as the corresponding observed rotation
curve. By fitting the \emph{simulated} rotation curve to the 
\emph{observed} rotation curve, we derived the intrinsic value of $\vm$.
Four examples of observed rotation curves
along with the best-fitting synthetic rotation curves are shown in 
Fig.~\ref{rcs}.

36 galaxies had to be rejected from the further analysis because
the $S/N$ was too low to probe the regime of constant rotation velocity
at large radii, or because the rotation curves were perturbated.
Four objects were reliable for the $\vm$ determination,
but their positions were located
outside the field--of--view of the HST/ACS mosaic imaging. 
34 objects had curves with a high degree of symmetry and clearly reached
into the ``flat'' regime,
we consider these as high quality data. 39 curves 
had a relatively small spatial extent or mild asymmetries,
these will be referred to as low quality data in the following. 
In total, our kinematic data set thus comprises 73 late--type galaxies 
at a mean look--back time of $\sim$\,5\,Gyr. The objects span the range
$25\,{\rm km/s} \le \vm \le 450\,{\rm km/s}$ with a median of 129\,km/s 
(high quality data only: $62\,{\rm km/s} \le \vm \le 410\,{\rm km/s}$ and
$\langle \vm \rangle = 154\,{\rm km/s}$).

\section{\label{scal}Scaling Relations at Intermediate Redshift}

\begin{figure}[t]
\centerline{
\hspace{-1cm}
\psfig{file=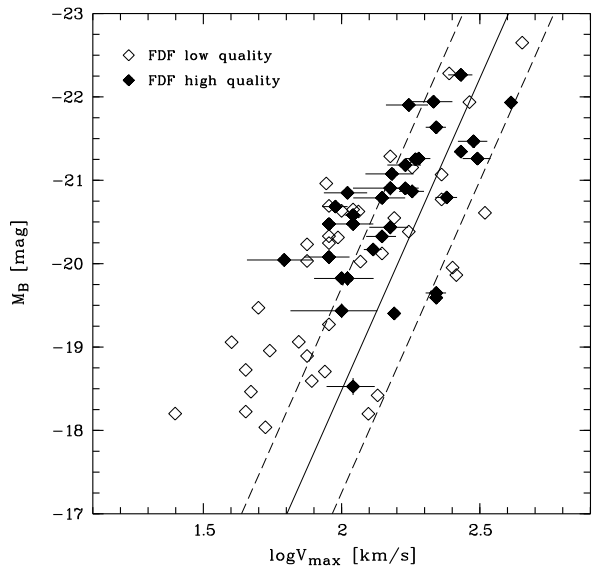,width=7.8cm,angle=-90}
}
\caption{\label{fdfpt}
FORS Deep Field sample of spirals in the range $0.1 \le z \le 1.0$ in 
comparison to the local Tully--Fisher relation  
by Pierce \& Tully (1992); the solid line denotes the fit to the local data,
the dashed lines 
give the 3\,$\sigma$ limits. The distant sample is
subdivided according to rotation curve quality: high quality curves (solid
symbols) extend well into the region of constant rotation velocity at 
large radii and therefore give robust values of $\vm$.
Error bars are shown for the high quality data only.
}
\end{figure}

In Fig.~\ref{fdfpt}, the maximum rotation velocities and absolute
magnitudes of the distant FDFTF galaxies are compared to the local
$B$-band Tully--Fisher relation by Pierce \& Tully (1992):
\begin{equation}
\label{local}M_B=-7.48\log \vm-3.52
\end{equation}
with a scatter of $\sigma_B=0.41^m$.
Note that, at variance with the original relation given by these authors,
we have calibrated the zero point to a face-on extinction of 0.27$^m$
to achieve consistency with the computation of the distant galaxies'
absolute magnitudes.
We emphasize that the further analysis is not sensitive to the choice
of the local reference sample: e.g., for the large data set of Haynes et 
al.~(1999, comprising 1097 objects), we find a very similar relation of
\begin{equation}
M_B=-7.85\log \vm-2.78,
\end{equation}
using a bisector fit (two geometrically combined least--square fits with the 
dependent and indepedent variable interchanged).
We will utilise the Pierce \& Tully sample here
for the sake of comparability to intermediate--redshift 
TF studies in the literature which 
mostly have used this sample as a local reference.

On the average, the distant galaxies are overluminous with respect to their
local counterparts, we find a median offset of 
$\langle \Delta M_B \rangle = -0.98^m$ for the total FDFTF sample and
$\langle \Delta M_B \rangle = -0.81^m$ for the high quality data only.
But we find also evidence for a differential evolution.
Fig.~\ref{fdfpt} indicates a relatively good agreement between the
intermediate--redshift
galaxies and the local TFR in the regime of fast rotators, i.e.~high masses,
while the distant low--mass galaxies systematically deviate from the relation
of present-day spirals. For low quality data, this may partly be due to
underestimated maximum rotation velocities, since the corresponding curves
have a relatively small spatial extent and do not robustly probe the region
of constant rotation velocity at large radii. 
In the case of high quality rotation curves, 
this is however unlikely, since these extent well into the ``flat region''.

A 100 iteration bootstrap bisector fit (average of 100 bisector fits with
randomly removed objects in each iteration) to the
34 FDFTF objects with high quality rotation curves yields
\begin{equation}
\label{bsq}M_B=-(4.05\pm0.58)\log \vm - (11.8\pm1.28),
\end{equation}
i.e.~the TFR slope we find at intermediate redshift is significantly shallower 
than in the local universe. Since the derivation of the galaxies' structural
parameters and of the $\vm$ values has been based entirely on HST/ACS imaging, 
Eq.~\ref{bsq} is a confirmation of the results presented in 
B\"ohm et al.~(2004) which were limited to ground--based imaging.

\begin{figure}[t]
\centerline{
\hspace{-1cm}
\psfig{file=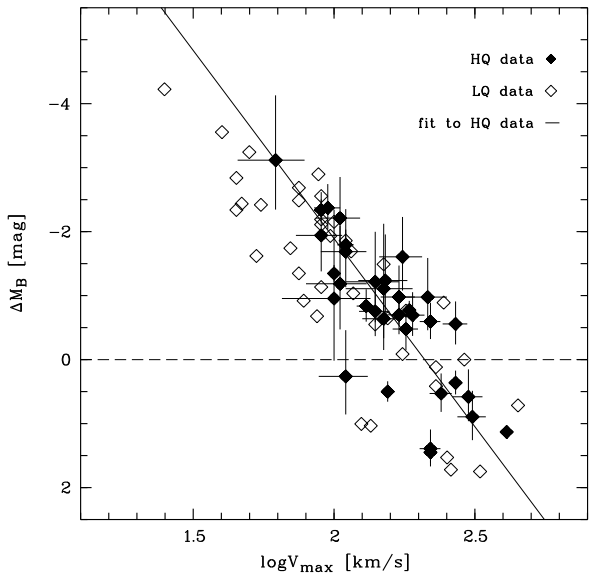,width=7.8cm,angle=-90}
}
\caption{\label{fdfpt2}
Offsets of the distant FORS Deep Field galaxies 
from the local TFR  by Pierce \& Tully (1992). The distant sample is
subdivided according to rotation curve quality: high quality curves (solid
symbols) extend well into the region of constant rotation velocity at 
large radii and therefore give robust values of $\vm$. The dashed horizontal line
corresponds a zero luminosity evolution. While high--mass galaxies are in
agreement with the local TFR or even slightly underluminous given their $\vm$,
the objects are increasingly overluminous towards small values of $\vm$
(error bars are shown for the high quality data only).
}
\end{figure}

We show the individual offsets $\Delta M_B$ of the FDF galaxies from
the local TFR as a function of their maximum rotation velocity in
Fig.~\ref{fdfpt2}. Even when restricting the sample to the high quality 
rotation curves, we find significant overluminosties of up to more than
2$^m$ in the rest-frame $B$ for low--mass spirals. $L^\ast$ galaxies,  
corresponding to $\log \vm \approx 2.3$ according to Eq.~\ref{local}, scatter
around a negligible evolution, while the fastest rotators are systematically
underluminous. 

\begin{figure}[t]
\centerline{
\hspace{-1cm}
\psfig{file=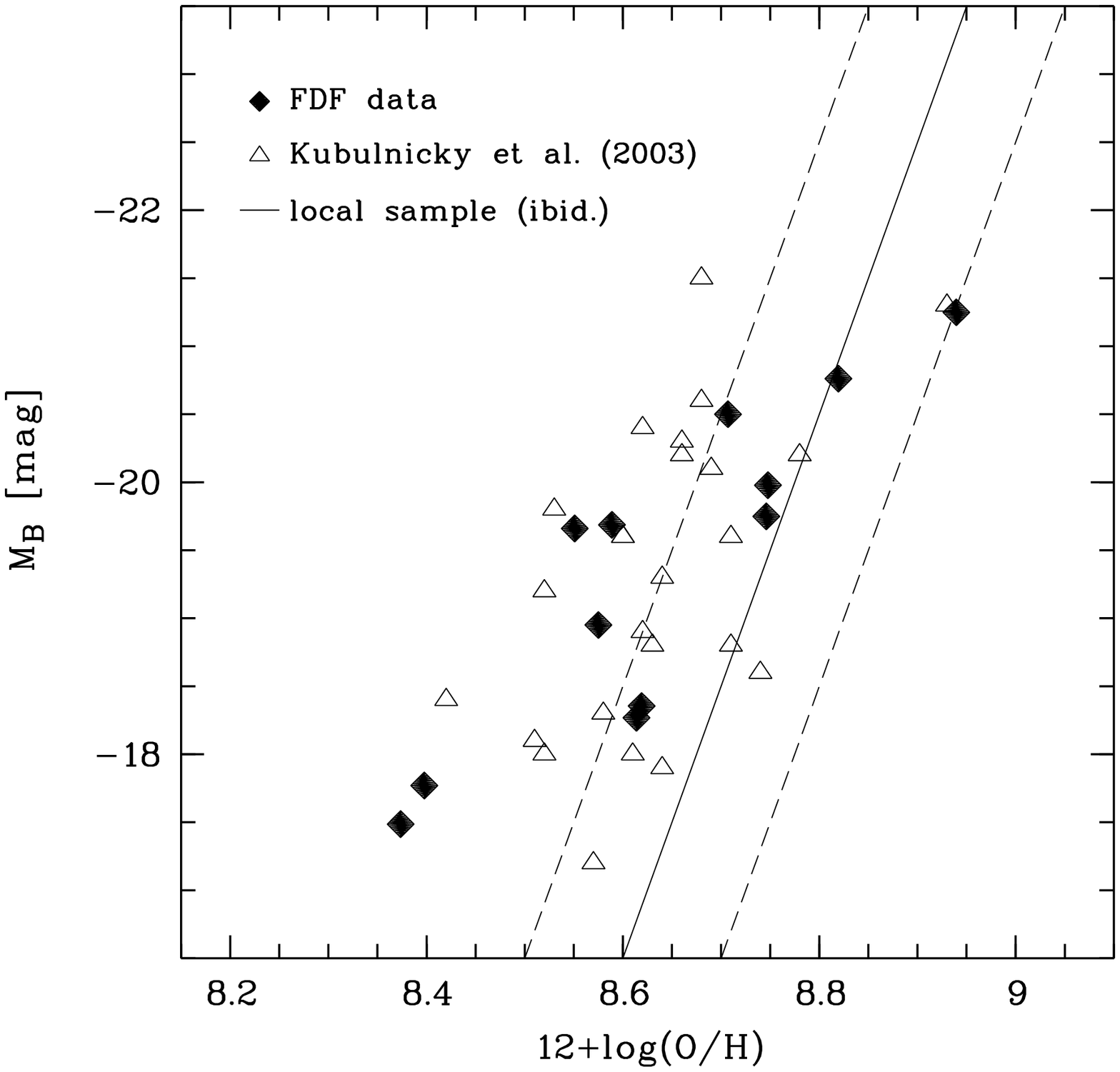,width=7.8cm,angle=-90}
}
\caption{\label{lz}
The gas-phase metallicities of 12 spirals from the FDF data set
(filled symbols) in comparison to the local luminosity-metallicity relation
constructed by Kobulnicky et al.~(2003,
solid line; dashed lines give the estimated 1$\sigma$ scatter). 
Also shown are distant galaxies  presented in
Kobulnicky et al.~(open symbols) which cover a similar redshift 
range as the FDF galaxies. 
Both distant samples show a ``tilt'' with respect to the local
relation which likelywise indicates a combined evolution in luminosity and
metallicity of low--luminosity galaxies.
}
\end{figure}

In Fig.~\ref{lz}, we show 
the sub-sample of 12 FDF galaxies for which we could determine the
oxygen abundances O/H in comparison to the local luminosity--metallicity 
relation as given in Kobulnicky et al.~(2003, the displayed scatter is a rough
estimate). In addition, a sub--sample
of distant late--type galaxies from the DEEP survey (ibid.) is shown,
which has been restricted to the same redshift intervall ($0.22<z<0.46$) 
that is covered by the FDF galaxies. 
Both sub--samples thus represent a look--back time of $\sim$\,4\,Gyr.
For the sake of comparability, the absolute magnitudes $M_B$ of the FDF
galaxies given in this figure are \emph{not} corrected for intrinsic absorption,
as is the case for the Kobulnicky et al.~data. 
Both distant samples indicate a  ``tilt'' with respect to the local
$L$--$Z$ relation. At given $\log {\rm (O/H)}$, high--metallicity galaxies
at intermediate redshift agree relatively well with the local
$L$--$Z$ relation, whereas low--metallicity objects are overluminous. 
Alternatively,
the distributions may be interpreted such that the distant low--luminosity
galaxies have smaller chemical yields than locally, while high--luminosity 
galaxies do not differ strongly in O/H between intermediate and low redshift.
If the offsets of the FDF spirals we observe in the TF diagram are due to 
younger stellar populations than locally, it is probable that Fig.~\ref{lz} shows
a combined evolution in luminosity \emph{and} metallicity. This indeed has
been the conclusion of Kobulnicky et al.~after comparison of their data to
single--zone models.

\begin{figure}[t]
\centerline{
\hspace{-1cm}
\psfig{file=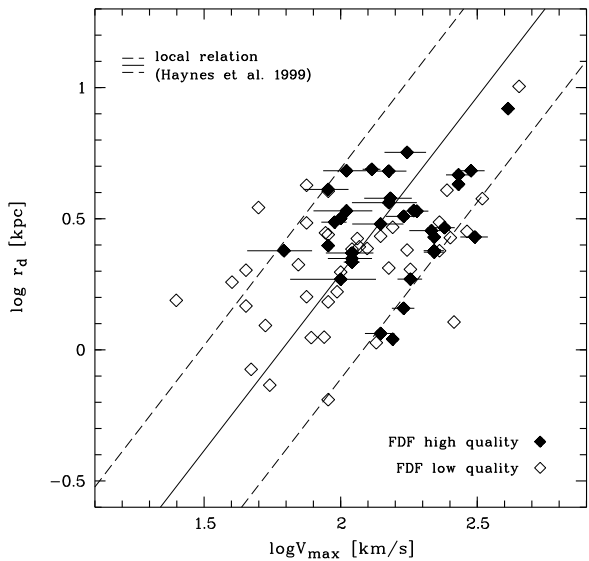,width=7.8cm,angle=-90}
}
\caption{\label{vsr}
Comparison between the intermediate--redshift FDF galaxies and the local
velocity--size relation of the Haynes et al.~sample (1999, solid line;
dashed lines correspond to the 3\,$\sigma$ scatter). 
Error bars are shown for high quality
data only.
}
\end{figure}

The third scaling relation which we want to focus on here is presented in
Fig.~\ref{vsr}, where the FDF galaxies are shown with respect to the local
velocity--size relation correlating $\vm$ and $\rd$. To derive the latter,
we used the sample of  Haynes et al.~(1999).
A bisector fit this data set yields
\begin{equation}
\log \rd = 1.35 \log \vm -2.41.
\end{equation}
Since the scale lengths of the FDF galaxies were determined in the $I$-band
(HST-filter F814W), the data intrinsically probe shorter wavelengths towards
higher redshifts. For direct comparability to the local sample ($I$-band 
as well),
we had to account for this rest--frame shift. Otherwise, the measured
FDF disk sizes could be overestimated, in particular for the more distant
galaxies. Adopting the relations between scale lengths of local spirals
at different wavelengths
presented in de Jong (1996), we have transformed the observer's frame $I$-band
scale lengths to the rest--frame $I$-band values. 
Note, however, that these factors
are relatively small: for the FDF galaxies at $z\approx1$ , the correction
corresponds to only $\sim$\,10\%.
The characteristic disk sizes of the FDFTF sample cover the range
$0.7\,{\rm kpc} \le \rd \le 10.1\,{\rm kpc}$ with a median 
$\langle \rd \rangle = 2.7\,{\rm kpc}$.

\section{\label{dis}Discussion}

The $\vm$-dependent TF offsets we observe at redshift
$\langle z \rangle \approx 0.5$ may be indicative for a significant
decrease of the luminosity of low--mass galaxies~--- 
possibly combined with an increase in metallicity~---
over the past $\sim$\,5\,Gyr
and a negligible evolution of high--mass galaxies.
This evolution would be at variance with theoretical
predictions: e.g., Steinmetz \& Navarro (1999) find mass--independent
TF offsets towards higher redshifts with an $N$-body Smoothed Particle
Hydrodynamics code. Boissier \& Prantzos (2001), who used
a ``backwards approach'' model calibrated to the observed
chemo--spectrophotometric
properties of local spirals, predict overluminosties of high--mass spirals
and underluminosities of low--mass spirals towards larger look--back times.

It has to be ruled out
that our result might be induced by an observational bias or a
systematic error. E.g., it is known that present--day spirals
show a correlation between their TF residuals and broad--band colors
(e.g.~Kannappan et al.~2002),
with blue galaxies preferentially populating the regime of
overluminosities. We have therefore tested whether
Fig.~\ref{fdfpt2} may simply reflect an evolution of the color--residual
relation with redshift, finding no evidence for such a trend
(B\"ohm et al.~2004).
Another issue that has to be adressed is the potential impact of sample
incompleteness.
Any magnitude--limited data set contains only a fraction of the objects
that are located within the observed volume. Towards lower luminosities
(or slower rotation velocities), this fraction becomes smaller. Furthermore,
the magnitude limit corresponds to higher luminosities at higher
redshifts. An incompleteness bias could therefore result in a flattening
of the distant TFR with increasing redshift. However, dividing our sample
into objects with $z \le 0.449$ (37 galaxies) and $z > 0.449$ (36 galaxies),
we find no evidence for such a redshift dependence, the respective slopes
of the two redshifts bins are $-3.49$ and $-3.77$. 
For a more sophisticated test of sample incompleteness
which is based on the work of Giovanelli et al.~(1997), we refer to
our results presented in Ziegler et al.~(2002).

In the following, we will address another three examples of tests we performed.
These are related to
the influence of the intrinsic rotation curve shape, the impact of the
intrinsic absorption correction and the issue of galaxy-galaxy interactions.
 
To derive the intrinsic maximum rotation velocity,
we have assumed an intrinsic rotation curve shape with a linear
rise of the rotation velocity at small radii which turns over into a
region of constant rotation velocity at a radius that depends on the
rest--frame wavelength of the used emission line. This shape is
observed for kinematically unperturbated, massive ($\sim L^\ast$) local
spirals (e.g., Sofue \& Rubin 2001). For galaxies of very high or
very low masses, on the other hand, it is observed
that even in the outer parts,
most rotation curves have a velocity gradient. While the rotation velocity
keeps rising beyond the ``turnover'' radius in very low--mass spirals,
the velocity gradient in very high--mass spirals is negative.
Persic et al.~(1996) have used $>$1000 curves of local spirals 
to derive a parameterisation that uses the luminosity of an object
as an indicator for the rotation curve shape. To ensure that the
observed TF offsets cannot be attributed to a false assumption on the
intrinsic rotational law, we have alternatively used this
so--called ``Universal Rotation Curve'' shape as input for the
computation of our synthetic velocity fields.
If we use the $\vm$ values derived this way to recompute the offsets from the
local TFR, the luminosity evolution we find is smaller by 
only $\sim$\,0.15$^m$ at $\vm \approx 100$\,km/s. Since this is  a 
very modest change of the offsets we found on the basis of the simple
``rise--turnover--flat''--shape 
(which have a median of $\langle \Delta M_B \rangle \approx -1.74^m$
at $\vm \approx 100$\,km/s for the HQ data), 
we conclude that our results do not differ
significantly between these two assumptions on the
intrinsic rotation curve shape.

Similarly, we have tested whether a different approach to correct for the
intrinsic absorption would have an effect on our results. All values
given here were derived following Tully \& Fouqu\'e (1985), i.e., the
amount of intrinsic absorption is assumed to depend
only on the inclination of
the disk. More recently,
Tully et al.~(1998) have found evidence that the dust reddening is~--
at least locally~---
stronger in high--mass spirals than in low--mass spirals. Using their
results, we have recomputed the absolute magnitudes of the FDFTF galaxies.
As a new local reference that is consistently corrected for intrinsic
absorption following Tully et al., we adopted the sample of Verheijen (2001)
which is slightly steeper (slope $-8.1$) than the Pierce \& Tully (1992) sample.
This is simply due to the fact that fast rotating, high--mass spirals
are assumed to have a larger amount of intrinsic absorption than in
the Tully \& Fouqu\'e approach, and vice versa in the low--mass regime. 
The offsets of the FDF high quality data
from the Verheijen TFR are however very similar to the initial values
($\langle \Delta M_B \rangle \approx -1.77^m$ vs.
$\langle \Delta M_B \rangle \approx -1.74^m$ at $\vm \approx 100$\,km/s).
With respect to the two conventions of intrinsic absorption correction
discussed here, the TF offsets therefore are robust.
 
A third aspect we want to focus on here concerns the interplay between
the stellar population properties and the environment. From studies in
the local universe, it is known that galaxies residing in close pairs
can be subject to tidal interactions which can increase the star formation
rates. In such cases, the fraction of high--mass stars would be enlarged
and, in turn, the mass--to--light ratio would be decreased, resulting in
overluminosities in the TF diagram. This triggering of star formation would
be particularly efficient in low--mass galaxies (e.g., Lambas et al.~2003).
Though we have selected our objects from a sky region that should be 
representative for low--density environments, it is not clear a priori
whether the 
correlation between the TF offsets
and $\vm$ can at least in part be attributed to tidally induced star formation. 
Based on $>$10$^5$ galaxies from the 2dF survey, Lambas et al.~have found that
the star formation rates
can be significantly increased in objects that have
close companions with a separation  
$\Delta V_{\rm sys} \le 250$\, km/s in systematic velocity
and a projected distance of
$D_{\rm proj} \le 100$\,kpc. Using all 267 available spectroscopic redshifts
of FDF galaxies at $z\le1$ 
from our own study and a low--resolution survey presented
in Noll et al.~(2004), 
and adopting the Lambas et al.~constraints cited above,
we have found 12 FDFTF
objects with confirmed neighbors. 

\begin{figure}[t]
\centerline{
\hspace{-1cm}
\psfig{file=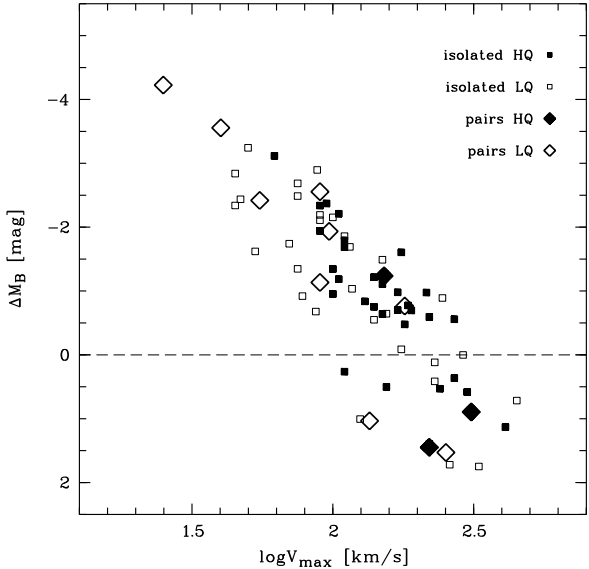,width=7.8cm,angle=-90}
}
\caption{\label{fdfpt4}
Offsets of the distant FORS Deep Field galaxies 
from the local TFR  by Pierce \& Tully (1992). Large symbols depict
FDF objects which show spectroscopically confirmed neighbors within
$\Delta V_{\rm sys} \le 250$\, km/s in systematic velocity
and a projected separation
$D_{\rm proj} \le 100$\,kpc. See text for details.
}
\end{figure}

In Fig.~\ref{fdfpt4},
we show the TF offsets of these galaxies in comparison to the rest of
the sample. Though the small sub--sample of pair candidates does not allow
robust statistics, the galaxies with close companions appear to be similarly 
distributed as the rest of the sample. Moreover,
we find a hint that the rotation curve quality is reduced with respect
to the probably isolated galaxies~--- only 3 out of 12 (25\%) pair candidates 
were
classified to have high quality rotation curves, whereas for the rest of the sample,
this fraction is 31 out of 61 (51\%). In particular, the pair candidates
with HQ curves are not systematically biased towards large overluminosities.
Since we have spectroscopic redshifts only for a fraction of the galaxies
within the probed volume, it is possible that we missed some close pairs. 
However, the aim was to test whether those pair candidates which \emph{are}
identified systematically differ from the remaining FDFTF objects.
We thus conclude that our analysis which is based on the high quality rotation 
curves is very unlikely to be affected by tidally induced star formation.

\begin{figure}[t]
\centerline{
\hspace{-1cm}
\psfig{file=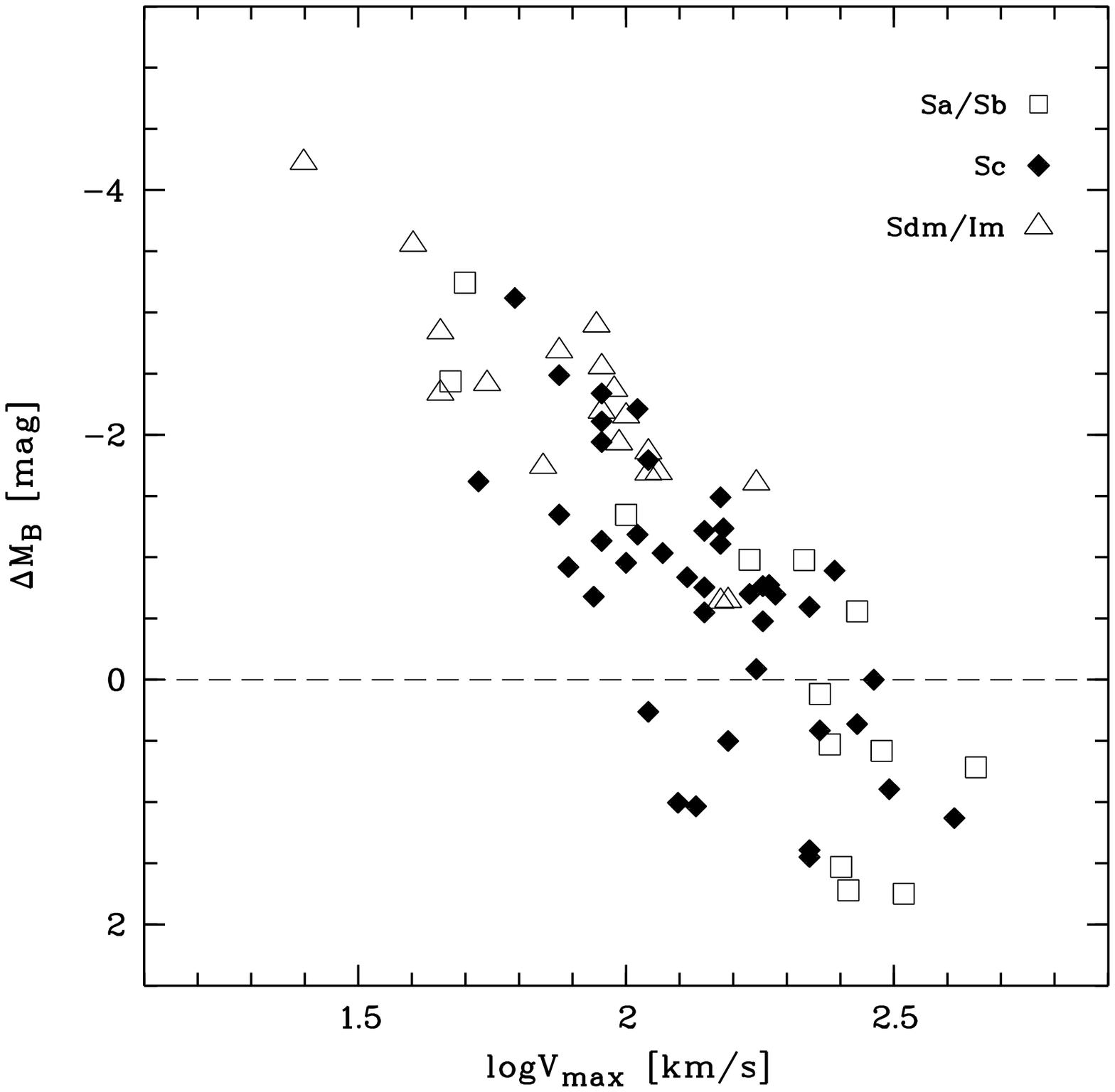,width=7.8cm,angle=-90}
}
\caption{\label{fdfpt3}
Offsets of the distant FORS Deep Field galaxies 
from the local TFR  by Pierce \& Tully (1992). The distant sample is
subdivided according to the SED type. All three sub--samples indicate
a $\Delta M_B$--$\vm$ correlation, but cover different mass regimes.
See text for details.
}
\end{figure}

To summarise all the tests performed, we find no evidence for any systematic
error or bias that may be the source of the observed
shallow slope of the intermediate--redshift TFR.
We now will show that our findings can be used to interpret the
rather discrepant results of previous TF studies of distant galaxies
introduced in Sect.~\ref{intro}.  
For this, Fig.~\ref{fdfpt3}
shows our sample sub--divided according to the Spectral Energy 
Distribution into early--type spirals (Sa/Sb), intermediate--types
(Sc) and very late--types (Sdm/Im). All three sub--samples
show a correlation between the TF offsets and $\vm$, but cover different
mass regimes: galaxies with late--type spectra have smaller average $\vm$ values
than early--type spirals. The respective classes have median values of
$\langle \vm \rangle_{\rm Sdm/Im} = 91$\,km/s, 
$\langle \vm \rangle_{\rm Sc} = 140$\,km/s and
$\langle \vm \rangle_{\rm Sa/Sb} = 240$\,km/s. For samples which are too small
to robustly test a correlation between $\Delta M_B$ and $\vm$ but can only
be used to determine \emph{average} TF offsets, this would results in a
correlation between SED type and $\langle \Delta M_B \rangle$.

Studies comprising only 10-20 galaxies which were selected on blue colors or 
strong emission lines and therefore
preferentially contained late--type spirals (e.g., Rix et al.~1997,
Simard \& Pritchet 1998), found evidence for large luminosity offsets at
intermediate redshift with respect to the local TFR. According to the
above, this can be attributed to a combination  of a selection
effect and small number statistics. Similary, a study by Vogt et al.~(1996)
which was selected on large disks and therefore mainly contained early-type
spirals (which on average are more massive than late--types), 
yielded a modest value for the mean TF offset. 

A more recent study on $\sim$100 intermediate--redshift spirals by Vogt
(2001) did not find evidence for a slope evolution of the TFR, 
at variance with our results. Moreover, this study yields only a modest
average TF offset of $-$0.2$^m$, compared to
the $-$0.8$^m$ we find for the HQ data (note that these values correspond
to the same cosmology and similar redshifts).
It is difficult to speculate whether this could be attributed to different
criteria of rotation curve quality or differences in the $\vm$ derivation
procedures. On the other hand, a sample of 64 galaxies drawn from the same
survey (the DEEP project, see Koo 2001) finds, in comparison to local galaxies, 
a tilt of the intermediate-redshift luminosity--metallicity relation which is 
also indicated by our data (Fig.~\ref{lz}).

Is is a complicated issue to determine the key processes which can
give rise to the mass--dependent TF offsets we observe.
Several effects likelywise act in combination when local and distant spirals 
are compared. E.g.,
the stellar populations of the intermediate--redshift galaxies
are probably younger than those of their local counterparts,
the gas mass fractions and chemical yields also evolve with time etc.
Since even a relatively small fraction of young, high--mass stars can have
a significant effect on the luminosity in the blue bands, a straightforward
interpretation would be that the TF offsets of the
FDF galaxies point towards a correlation between mass and
\emph{age}.

As a first deeper insight into this matter, Ferreras et al.~(2004) have
used single--zone models of chemical enrichment on a  sub--sample of the 
FDFTF galaxies at $z>0.5$. 
These models were determined by only four free parameters:
formation redshift, gas infall timescale, star formation
efficiency and gas outflow fraction.
Model star formation histories were generated which, combined with the latest 
Bruzual \& Charlot models (2003), were used to compute simulated 
$UBgRIJK$ broad-band colors. Probing a large volume in parameter space,
these synthetic colors were fitted
to the observed broad-band
colors, thus deriving the four model parameters for each of the
$z>0.5$ FDF galaxies. The best--fitting models indicated 
that high--mass galaxies on the average have higher star formation
efficiencies, with a ``break'' at $\vm \approx 140$\,km/s which~---
for reasonable $M/L$ ratios~--- interestingly is good agreement with
the results found by Kauffmann et al.~(2003) for \emph{local} galaxies
from the Sloan Digital Sky Survey.

Moreover, the best--fitting formation redshifts of the Ferreras et al.~models
were found to be higher for the more massive FDFTF galaxies than for low--mass
galaxies. The models hence yielded evidence that high--mass spirals
started to convert their gas into stars at earlier cosmic epochs and 
on shorter timescales than low--mass ones. When evolved to zero redshift,
the mean model stellar ages turned out to be 
older in high--mass galaxies than in low--mass
galaxies. These results hint towards an \emph{anti--hierarchical evolution}
of the stellar (baryonic) component, a phenomenon 
that recently has been 
referred to as ``down-sizing'' (e.g., Kodama et al.~2004).

\begin{figure}[t]
\centerline{
\hspace{-1cm}
\psfig{file=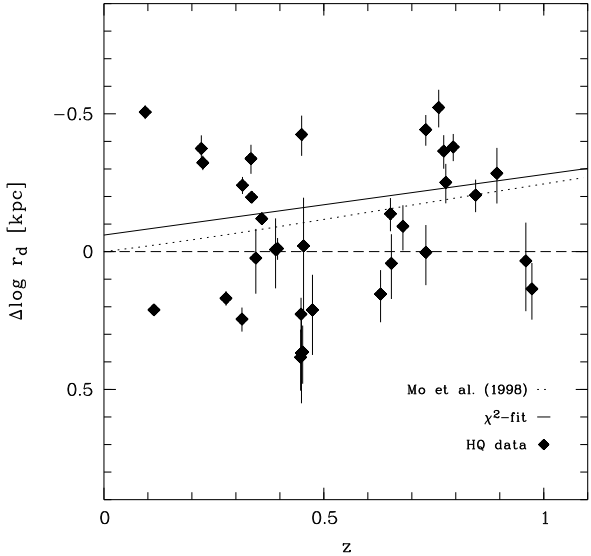,width=7.8cm,angle=-90}
}
\caption{\label{fdfvsr}
Offsets $\Delta \log \rd$
of the distant FORS Deep Field galaxies (high quality rotation curves
only) from the local velocity--size relation of the sample by Haynes et 
al.~(1999) shown in Fig.~\ref{vsr}, plotted as a function of redshift. 
Objects with $\Delta \log r_{\rm d}>0$ have larger disks than local spirals at 
a given $\vm$, whereas values $\Delta \log r_{\rm d}<0$ correspond to disks 
which are smaller than in the local universe. As indicated by the fit
to the data (solid line), we find a slight trend towards smaller disks at
higher redshifts, in agreement with theoretical predictions 
within the hierarchical scenario (Mo et al.~1998, dotted line).
}
\end{figure}

It is however improbable that this implies a contradiction to the 
hierarchical growth of the Cold Dark Matter halos. In Fig.~\ref{fdfvsr},
we show the offsets $\Delta \log \rd$
of the FDFTF galaxies with high quality rotation curves from
the local velocity--size relation (which we presented in Fig.~\ref{vsr})
as a function of \emph{redshift}.
Given their maximum rotation velocity,
distant galaxies with $\Delta \log \rd<0$ have
disks that are smaller than in the local universe,
while the disks of galaxies with $\Delta \log \rd>0$ are larger.
Though the scatter in $\Delta \log \rd$
is substantial, we observe a slight trend towards smaller disk sizes
at higher redshifts. This is in compliance with the results from other
observational studies (e.g., Giallongo et al.~2000, Ferguson et al.~2004).
Moreover, as is depicted in Fig.~\ref{fdfvsr},
the fit to the data is in relatively good agreement with the
prediction of disk growth in the hierarchical scenario (e.g. Mo et al.~1998).

On the one hand, we find that the observed luminosity evolution of the FDF 
galaxies deviates from the results of simulations which were used to
predict the TFR of distant spirals. On the other hand, the intermediate--redshift
disks are observed to be smaller than locally, which is in compliance with
the CDM hierarchical scenario. The results from single--zone models
might indicate that the star formation is suppressed in low--mass galaxies
due to, e.g., SN feedback (cf.~Dalcanton et al.~2004). Potentially,
the discrepancies between observations and simulations arise from the fact
that the mechanisms suppressing the star formation are not yet implemented 
realistically enough in models of galaxy evolution. 
 
\section{\label{concl}Conclusions}

Using the FORS instruments of the VLT in 
multi--object spectroscopy mode and HST/ACS imaging, 
we have observed a sample 113 disk galaxies 
in the FORS Deep Field. The galaxies reside at redshifts 
$0.1<z<1.0$ and thereby
probe field galaxy evolution over more than half the age of the universe.
All spectrophotometric types from Sa to Sdm/Irr are comprised.
Spatially resolved rotation curves have been extracted and fitted with
synthetic velocity fields that account for geometric distortions as well
as blurring effects arising from seeing and optical beam smearing.
The intrinsic maximum rotation velocities $V_{\rm max}$ were derived for
73 galaxies within the field--of--view of the ACS images. 
Two--dimensional surface brightness profile fits were performed to measure
the structural parameters like disk inclinations, position angles etc.

The massive distant galaxies fall onto the local Tully--Fisher Relation, while 
the low--mass distant galaxies are brighter than locally by up to $>$2$^m$ in 
rest--frame $B$. This trend might be combined with an evolution in 
metallicity. We find no evidence for a bias or 
systematic errors that could induce the observed shallow slope of the
Tully--Fisher Relation at intermediate redshifts.
Discrepancies between several previous studies could be explained
as a combination of selection effects and small number statistics
on the basis of such a mass--dependent luminosity evolution.
On the other hand, this evolution would be at variance with 
the predictions from numerical simulations. 
For a given $\vm$, the disks of the distant 
galaxies are slightly smaller than those of their local counterparts, 
as expected for a hierarchical structure growth. 
Our results therefore are discrepant with theoretical predictions
only in terms of the stellar populations properties.
A possible explanation could be the suppression of star formation
in low--mass disks which is not
yet properly implemented in models of galaxy evolution. \\[0.5cm]
{\small 
{\bf  Acknowledgements} \\
This study is based on observations with the European Southern
Observatory Very Large Telescope 
(observing run IDs 65.O-0049, 66.A-0547 and 68.A-0013).
We are grateful 
for the continuous support of our project by the PI of the FDF
consortium, Prof.~I.~Ap\-pen\-zel\-ler (LSW Heidelberg),
and by Prof.~K.~J.~Fricke (USW G\"ottingen).
We also thank Drs.~J.~Heidt (LSW Heidelberg), 
D.~Mehlert (LSW Heidelberg) and S.~Noll (USW M\"unchen)
for performing the
spectroscopic observations and ESO for the efficient support during the
observations.
Furthermore we want to thank J.~Fliri and
A.~Riffeser (both USW M\"unchen) for the cosmic ray rejection on the ACS images
and Dr.~B. Milvang-Jensen and M.~Panella (both MPE Garching)
for fruitful discussions.
Our work was funded by the Volkswagen Foundation (I/76\,520) 
and the Deutsches Zentrum f\"ur Luft- und Raumfahrt (50\,OR\,0301).
}

\subsection*{References}

{\small

\bref
Bender, R., Appenzeller, I., B\"ohm, A., et al. 2001, 
The FORS Deep Field: Photometric redshifts and object classification,
in Deep Fields, ed. S.~Cristiani, A.~Renzini, \& R.~E.~Williams, 
ESO astrophysics symposia (Springer), 96

\bref
Bertin, E., \& Arnouts, S. 1996, A\&AS, 117, 393

\bref
Boissier, S., \& Prantzos, N. 2001, MNRAS, 325, 321

\bref
B\"ohm, A., Ziegler, B.~L., Saglia, R.~P., et al. 2004, A\&A, 420, 97

\bref
Bruzual, A. G., \& Charlot, S. 2003, MNRAS, 344, 100

\bref
de Jong, R. S. 1996, A\&A, 313, 377

\bref
Dalcanton, J. J., Yoachim, P., \& Bernstein, R. A. 2004, ApJ, 608, 189

\bref
Ferguson, H. C. F., Dickinson, M., Giavalisco, M., et al. 2004, ApJ, 600, L107

\bref
Ferreras, I., \& Silk, J. 2001, ApJ, 557, 165

\bref
Ferreras, I., \& Silk, J., B\"ohm, A., \& Ziegler, B. L. 2004, MNRAS, 355, 64

\bref
Giallongo, E., Menci, N., Poli, F., et al. 2000 ApJ, 530, L73

\bref
Giovanelli, R., Haynes, M. P., Herter, T., et al. 1997, AJ, 113, 53

\bref
Governato, F., Mayer, L., Wadsley, J., et al. 2004, ApJ, 607, 688

\bref
Haynes, M. P., Giovanelli, R., Chamaraux, P., et al. 1999, AJ, 117, 2039

\bref
Heidt, J., Appenzeller, I., Gabasch, A., et al. 2003, A\&A, 398, 49 

\bref
Kannappan, S. J., Fabricant, D. G., \& Franx, M. 2002, AJ, 123, 2358

\bref
Kauffmann, G., Heckman, T. M., White, S. D. M., et al. 2003, MNRAS, 341, 54
	 
\bref
Kobulnicky, H. A., Willmer, C. N. A., Phillips, A. C., et al. 2003, 
ApJ, 599, 1006

\bref
Kodama, T., Yamada, T., Akiyama, M., et al. 2004, MNRAS, 350, 1005

\bref
Koo, D. C. 2001, DEEP: Pre--DEIMOS Surveys to I$\sim$\,24 of Galaxy Evolution 
and Kinematics, in Deep Fields, ed. S.~Cristiani, A.~Renzini, 
\& R.~E.~Williams, ESO astrophysics symposia (Springer), 107

\bref
Lambas, D. G., Tissera, P. B., Sol Alonso, M., \& Coldwell, G. 2003, 
MNRAS, 346, 1189

\bref
Mathewson, D. S., \& Ford, V. L. 1996, ApJS, 107, 97

\bref
McGaugh, S. 1991, ApJ, 380, 140

\bref
Milvang-Jensen, B., Arag\'on-Salamanca, A., Hau, G. K. T., 
J{\o}rgensen, I., \& Hjorth, J. 2003, MNRAS, 339, L1

\bref
Mo, H. J., Mao, S., \& White, S. D. M. 1998, MNRAS, 295, 319

\bref
Navarro, J. F., \& White, S. D. M. 1994, MNRAS, 267, 401

\bref
Noll, S., Mehlert, D., Appenzeller, I., et al. 2004, A\&A, 418, 885

\bref
Peacock, J. A. 2003, RSPTA, 3671, 2479

\bref
Peng, C. Y., Ho, L. C., Impey, C. D., \& Rix, H.-W. 2002, AJ, 124, 266

\bref
Persic, M., \& Salucci, P., \& Stel, F. 1996, MNARS, 281, 27

\bref
Pierce, M. J., \& Tully, R. B. 1992, ApJ, 387, 47

\bref
Rix, H.-W., Guhathakurta, P., Colless, M., \& Ing, K. 1997, MNRAS, 285, 779

\bref
Sakai, S., Mould, J. R., Hughes, S. M. G., et al. 2000, ApJ, 529, 698

\bref
Simard, L., \& Pritchet, C. J. 1998, ApJ, 505, 96

\bref
Sofue, Y. \& Rubin, V. 2001, ARA\&A, 39, 137

\bref
Spergel, D. N., Verde, L., Peiris, H. V., et al. 2003, ApJS, 148, 175

\bref
Steinmetz, M., \& Navarro, J. F. 1999, ApJ, 513, 555

\bref
Tully, R. B., \& Fisher, J. R. 1977, A\&A, 54, 661

\bref
Tully, R. B., \& Fouqu\'e, P. 1985, ApJS, 58, 67

\bref
Tully, R. B., Pierce, M. J., Huang, J.-S., et al. 1998, AJ, 115, 2264

\bref
van den Bosch, F. 2002, MNRAS, 332, 456

\bref
Verheijen, M. A. W. 2001, ApJ, 563, 694

\bref
Vogt, N. P., Forbes, D. A., Phillips, A. C., et al. 1996, ApJ, 465, L15

\bref
Vogt, N. P., Phillips, A. C., Faber, S. M., et al. 1997, ApJ, 479, L121

\bref
Vogt, N. P. 2001, 
Distant Disk Galaxies: Kinematics and Evolution to z$\sim$1,
in Deep Fields, ed. S.~Cristiani, A.~Renzini, \& R.~E.~Williams, 
ESO astrophysics symposia (Springer), 112

\bref
Ziegler, B. L., B\"ohm, A., Fricke, K. J., et al. 2002, ApJ, 564, L69
}

\vfill

\end{document}